\pdfoutput=1
\RequirePackage{fix-cm}
\documentclass[twocolumn]{svjour3}          % twocolumn
\smartqed  % flush right qed marks, e.g. at end of proof
\usepackage{graphicx}
\usepackage{color}% Include figure files
\usepackage{dcolumn}% Align table columns on decimal point
\usepackage{amsmath,amssymb}% AMS
\usepackage{bm}% bold math
\journalname{}
\begin{document}

\title{Improving sequencing by tunneling with multiplexing and cross-correlations%\thanks{Grants or other notes
}


\author{P. Boynton         \and
        A. V. Balatsky     \and
        I. K. Schuller     \and
        M. Di Ventra %etc.
}


\institute{P. Boynton \at
              Department of Physics, University of California, San Diego, La Jolla, CA 92093-0319 \\
              \email{pboynton@physics.ucsd.edu}           %  \\
           \and
           A. V. Balatsky \at
              Nordita Roslagstullsbacken 23, 106 91 Stockholm Sweden
              \at
              Institute for Materials Science, Los Alamos, NM 87545
           \and
           I. K. Schuller \at
              Department of Physics, University of California, San Diego, La Jolla, CA 92093-0319
           \and
           M. Di Ventra \at
              Department of Physics, University of California, San Diego, La Jolla, CA 92093-0319
}

\date{Received: date / Accepted: date}

\maketitle

\begin{abstract}
%150 to 250 words
Sequencing by tunneling is a next-generation approach to read single-base information using electronic tunneling transverse to the single-stranded DNA (ssDNA) backbone while the latter is translocated through a narrow channel. The original idea considered a single pair of electrodes to read out the current and distinguish the bases \cite{zwolak2005electronic,Lagerqvist2006}. Here, we  propose an improvement to the original sequencing by tunneling method, in which $N$ pairs of electrodes are built in series along a synthetic nanochannel. While the ssDNA is forced through the channel using a longitudinal field it passes by each pair of electrodes for long enough time to gather a minimum of $m$ tunneling current measurements, where $m$ is determined by the level of sequencing error desired. Each current time series for each nucleobase is then cross-correlated together, from which the DNA bases can be distinguished. We show using random sampling of data from classical molecular dynamics, that indeed the sequencing error is significantly reduced as the number of pairs of electrodes, $N$, increases. Compared to the sequencing ability of a single pair of electrodes, cross-correlating $N$ pairs of electrodes is {\it exponentially} better due to the approximate log-normal nature of the tunneling current probability distributions. We have also used the Fenton-Wilkinson approximation to analytically describe the mean and variance of the cross-correlations that are used to distinguish the DNA bases. The method we suggest is particularly useful when the measurement bandwidth is limited, allowing a smaller electrode gap residence time while still promising to consistently identify the DNA bases correctly.
%4 to 6 keywords
\end{abstract}
%Don't forget to give each section and subsection a unique label.
%\paragraph{Paragraph headings} Use paragraph headings as needed.
%
\section{Introduction}
\label{sec:intro}

A cheap and fast method to sequence DNA would revolutionize the way health care is conducted \cite{Zwolak2008}. With such a method, medicine would be catered to the individual based on genetic implications, an approach that goes under the name of personalized or precision medicine. The research behind DNA sequencing is rich and plentiful, with many techniques that have much potential. Two of the most successful techniques currently used, single molecule real time sequencing (SMRT) \cite{eid2009real} and ion torrent semiconductor sequencing (ITS) \cite{rusk2010torrents}, need on the order of 10 hours, including full preparation time, for one run, which sequences 1 Gb and 100 Mb, respectively \cite{quail2012tale}. Both techniques take advantage of massively-parallel sequencing to achieve these benchmarks.

However, most of the current sequencing techniques, SMRT included, require fluorescent dyes to distinguish the DNA bases \cite{quail2012tale}. In other words, these techniques cannot greatly improve in speed and are inherently costly, both for the sample preparation, equipment and to operate. On the other hand, ITS does not utilize fluorescent dyes but instead depends on the detection of hydrogen ions released once a deoxyribonucleotide triphosphate (dNTP) forms a covalent bond with a complementary nucleotide \cite{rusk2010torrents}. This means that the overall costs are smaller in comparison but the technique nevertheless suffers from small read lengths of about 200 base pairs per run \cite{quail2012tale}, implying the technique would be difficult (or too costly) to apply to {\it de novo} sequencing.

\begin{figure}
\includegraphics[width=\columnwidth]{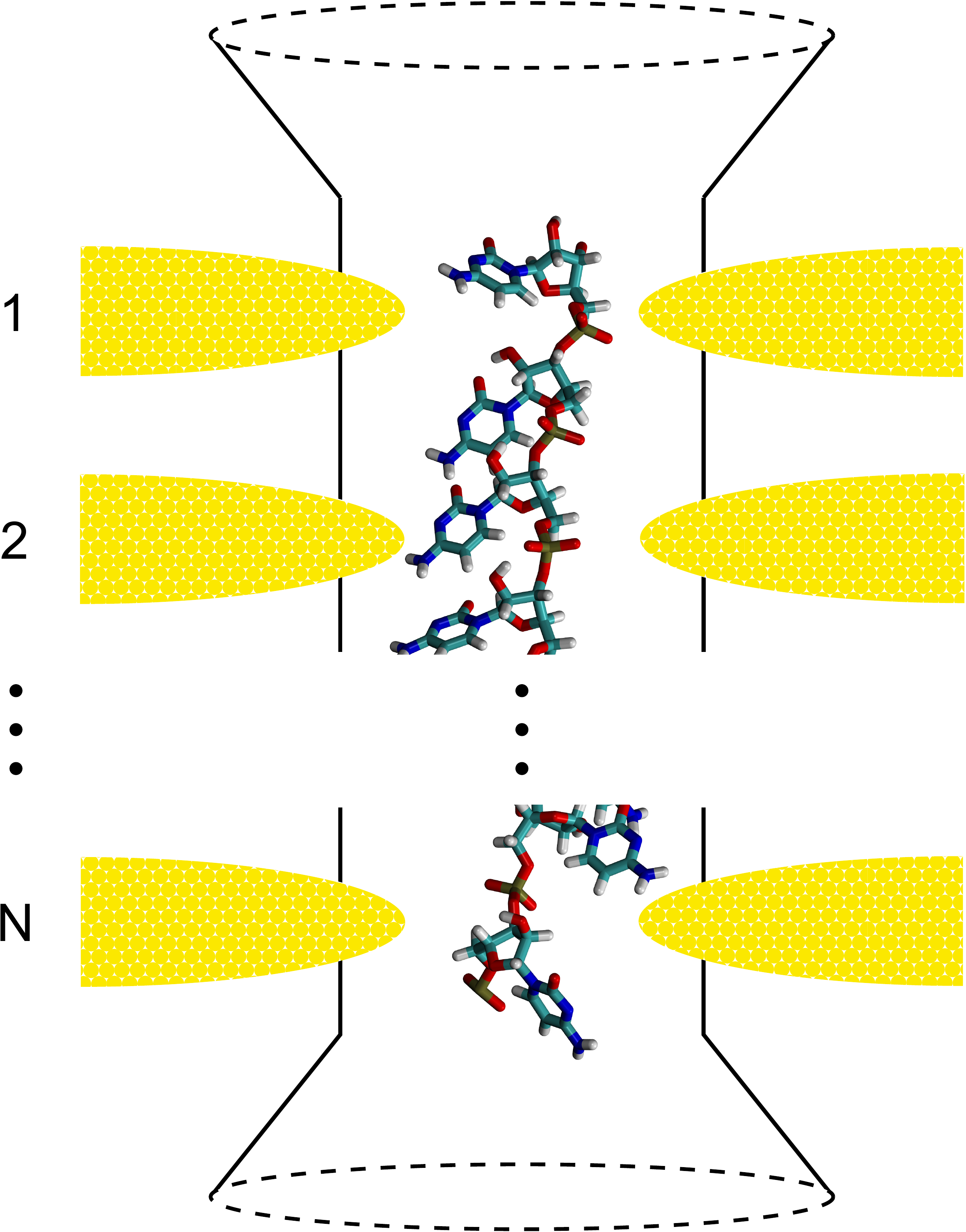}
\caption{\label{fig:setup}(Color online) Schematic for the multiplexed transverse electronic sequencing device. The solid-state nanopore is outlined in black with the dashed lines representing the conical entrance/exit that leads to the cylindrical nanochannel. Within the nanochannel is the ssDNA strand to be sequenced and several pairs of embedded gold electrodes labeled 1, 2, ... N to indicate the existence of a series of $N$ pairs of electrodes. Each pair of electrodes would be attached to a voltage source so that the DNA bases align with the field and the tunneling current flows stronger. In addition, a single pair of biased electrodes would be placed diametrically opposite above and below the pore to push/pull the negatively charged ssDNA strand through the nanochannel.}
\end{figure}

Quite recently a new approach has been suggested that envisions the sequencing of single-stranded DNA (ssDNA) with electronic currents transverse to the DNA backbone as it passes through a nanochannel \cite{zwolak2005electronic,Lagerqvist2006}. A schematic is shown in Fig. \ref{fig:setup}. This approach has been recently demonstrated experimentally by sequencing micro-RNA and short DNA oligomers \cite{ohshiro2012single}.

When the electrodes are fabricated so that the gap only allows a single base to fit at a time \cite{Tsutsui2008}, one can truly obtain single-base discrimination without the need of amplification or chemicals. Because of the speed of electronic-based detection, one can achieve sequencing rates of 1.2 Mb/hour with 0.1\% error per base without accounting for any parallelism or preparation time. This rate can be achieved with only 10 kHz sampling rate \cite{tsutsui2011electrical}, given that about 30 measurements are needed per base (derived using data from \cite{Krems2009}). An increase of sampling rate to 1 MHz would achieve a sequencing rate of 120 Mb/hour with the same error. Finally, increasing the error by an order of magnitude would only slightly decrease the sequencing rate \cite{Lagerqvist2006}. Since this nanopore method does not require the ssDNA strand to be of a certain length to function, the read length depends solely on the sequencing device's bandwidth and its ability to keep the ssDNA strand untangled and consistently translocating through the pore. In addition, as a label-free method, the technique benefits from a modest preparation time and reduced operating costs.

On the other hand, due to the speed of translocation of the ssDNA and the linear width of a single nucleotide, roughly 6.3 {\AA} \cite{murphy2004probing}, the current through each nucleotide has to be measured in a short period of time with limited bandwidth. Experiments have found the translocation speed to be difficult to control \cite{tsutsui2011electrical,Tsutsui2010}, yet the gate modulation of nanopore surface charges promises to reduce this speed and add an element of control to the instantaneous velocity of the ss-DNA strand \cite{he2011controlling}. With few current measurements per base, it becomes hard to identify the sequence of the ss-DNA strand without substantial errors. Therefore, if bandwidth is an issue, we suggest the use of a nanochannel containing several pairs of electrodes in series like in a multiplexing configuration, as shown in Fig. \ref{fig:setup}. We show that the signals from each pair of electrodes can be cross-correlated to significantly reduce noise and consequently reduce errors in base identification. To prove this point we have analyzed the cross-correlations of many ssDNA translocation realizations, finding that with a limited bandwidth already two pairs of electrodes far surpass the sequencing capability offered by a single pair. The approach we propose expands upon the recent work by Ahmed {\it et al.} \cite{ahmed2014correlation}, where the multiple electrode current readout was considered for the case of a multilayer graphene nanopore \cite{garaj2010graphene}. Here, we use the molecular dynamics simulations to characterize the noise along with a different cross-correlation analysis to estimate the signal to noise improvements on the multiple contact readout of solid-state nanochannels.

In experiments, the signal from electrons tunneling through a single nucleotide of ssDNA switches between a high average current state to a low average background current state in a pulse-like manner \cite{tsutsui2011electrical,Tsutsui2010,Huang2010}. Short episodes of background current occur because of the changing adsorption between the DNA base and the electrodes while long episodes are explained by the absence of a DNA base. Using current thresholds and the time spent in each background current episode one can mark the beginning and end of each nucleotide in the time series. With this method the $j$th nucleotide that travels through the first electrode pair can be matched with the $j$th nucleotide that travels through the following electrode pairs for cross-correlation. After the current time series from the $i$th electrode pair for the $j$th nucleotide is isolated, the short episodes of background current can be removed to leave only the pulses of current indicative of tunneling through the $j$th nucleotide of ssDNA. We define the resultant signal as $\mathcal{I}_{i}^{j}$.
\section{Molecular Dynamics Methods}
\label{sec:md_methods}

To simulate this process, we first use a combination of molecular dynamics (MD) performed with NAMD2 \cite{Phillips2005} and quantum transport calculations to obtain a current time series from a single electrode pair for each of the four bases: adenine ($A$), cytosine ($C$), guanine ($G$), and thymine ($T$). The contributions from neighboring nucleotides to the current have been found to be negligible provided the electrode cross-section is on the order of 1 nm \cite{zwolak2005electronic}. The MD results we use here have been taken from previous work in \cite{Krems2009} where the simulation proceeds as follows. A double-conical ${\rm Si}_3 {\rm N}_4$ nanopore with embedded gold electrodes in the center is built with a minimum diameter of 1.4 nm and a maximum diameter of 2.5 nm (similar to Fig. \ref{fig:setup} with just one electrode pair). The inner diameter is such that the homogeneous ssDNA can just pass through so that the electrode spacing can be at a minimum to enhance the signal. The ssDNA is placed parallel to the longitudinal axis so that the first base has past the entrance of the pore. The pore-DNA system is solvated in a TIP3P water sphere and constrained with periodic boundary conditions in an NVT ensemble with a 1 M solution of ${\rm K}^+$ and ${\rm Cl}^-$. The system is evolved in time with 1 fs steps and kept at room temperature with Langevin damping. To drive the ssDNA through the pore within a feasible simulation time a global longitudinal electric field of 6 ${\rm kcal} / ( {\rm mol} \, {\mbox \AA} \,e )$ is applied. When a base of ssDNA sits in between the electrodes the longitudinal pulling field is turned off and a transverse field of the same magnitude is turned on to calculate the electronic transport. This is an approximation to the transverse field being much larger than the longitudinal field, which is the optimum operating regime for the present sequencing device as the bases are better aligned with the transverse field \cite{Lagerqvist2006}.

The current is calculated with a single-particle elastic scattering approach using a tight-binding Hamiltonian \cite{DiVentra2008}. Coordinate snapshots of the molecular dynamics are taken every ps, with which a tight-binding Hamiltonian is created for the region between the gold electrodes. The Fermi level is taken to be that of bulk gold. To obtain the tunneling current through the ssDNA, we use the single-particle retarded Green's function,
\begin{equation}
G_{\rm DNA}(E) = \frac{1}{E S_{\rm DNA} - H_{\rm DNA} - \Sigma_t - \Sigma_b},
\label{eq:green}
\end{equation}
where $E$ is the energy, $S_{\rm DNA}$ and $H_{\rm DNA}$ are the overlap and Hamiltonian matrices, respectively, of the electronic junction, and $\Sigma_t$ and $\Sigma_b$ are the top and bottom electrode self-energies, respectively, for the interaction with the junction contents. The Green's function for gold needed to calculate $\Sigma_t$ and $\Sigma_b$ is approximated as in \cite{Pecchia2003}. The transmission function is obtained from the Green's function and the self-energies in the usual way (see, e.g., \cite{DiVentra2008}). The current is then given by
\begin{equation}
I = \frac{2 e}{h} \int^{\infty}_{-\infty}{dE\,T(E)[f_t(E) - f_b(E)]},
\label{eq:current}
\end{equation}
where $e$ is the elementary charge, $h$ is Planck's constant, $E$ is the energy of the scattering electron, $T$ is the total transmission function, and $f_t$ and $f_b$ are the top and bottom electrode Fermi-Dirac distribution functions, respectively \cite{DiVentra2008}. This process is carried out for every snapshot to obtain a time series for each of the four bases.
\begin{figure}
\includegraphics[width=\columnwidth]{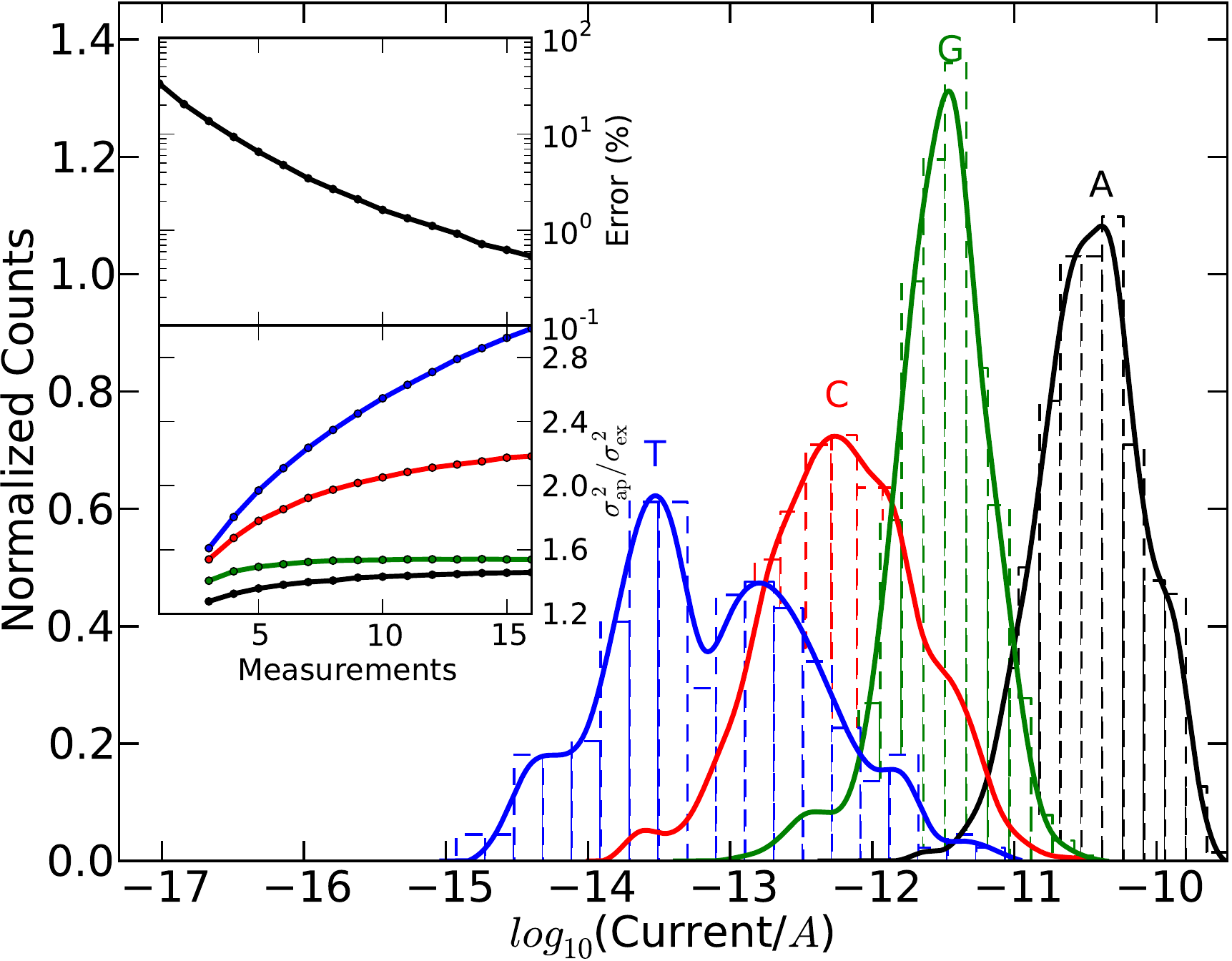}
\caption{\label{fig:one_pair}(Color online) Normalized current distributions from \cite{Krems2009} for the four bases, $A, C, G, T$, with one pair of electrodes, where the solid lines are cubic spline mirror-symmetric interpolations of the dashed line histograms. The distributions describe the probability of the base-10 log of the current due to the multi-scale nature of tunneling currents. The upper inset plots the sequencing error percentage per base on a log scale against the number of measurements per base, $m$. The lower inset plots the Fenton-Wilkinson approximated variance (see Eq. \eqref{eq:fenton_var}) for $\sigma_{N = 2,m}^{2}$ divided by the exact variance of ${\rm log}(g_{N = 2,m}^{j})$ against $m$ for $j = A, C, G, T$, where the color of the line corresponds to the base whose distribution has the same color.}
\end{figure}

The points in the time series are, to a good approximation, independent since the time for electrons to tunnel (${\sim}10^{-15}$ s) is much smaller than the time between each snapshot recording ($10^{-12}$ s). This coincides with experiments where we expect each point in $\mathcal{I}_{i}^{j}$ to be effectively independent since the time scale governing the molecular disorder that modulates the current (the fastest being water at ${\sim}10^{-12}$ s \cite{Mitra2001}) is much smaller than the typical time scale of measurement (${>}10^{-6}$ s or a kHz sampling rate \cite{tsutsui2011electrical,Tsutsui2010,Huang2010}). As a result, we do not expect the cross correlations to cut out these fast noise time scales in the current, rather the slowly propagating modes.

A probability distribution for the current values is created for each of the four bases by binning each respective time series as seen in Fig. \ref{fig:one_pair}. From these probability distributions we construct a set of time series, $\{I_{i}^{j}\}$, that resemble the signals generated by a ssDNA passing through a nanochannel with $N$ pairs of electrodes, or $\{\mathcal{I}_{i}^{j}\}$. Each $I_{i}^{j}$ is the tunneling current time series from the $i$th electrode pair and the $j$th nucleotide in the ssDNA that is centered around $t=0$ for convenience. Given that the spacing between opposing electrodes is roughly equivalent from electrode pair to electrode pair along the nanochannel, the pore-electrode environment would be nearly identical in each case.
\section{Cross-correlations}
\label{sec:cross}

Due to the independence of $\mathcal{I}_{i}^{j}$ and $I_{i}^{j}$ we can use a Monte Carlo method in which numbers are generated from a uniform distribution and then matched to a current value in the cumulative distribution function (cdf) for the $j$th nucleotide to create the set of $\{I_{i}^{j}\}$. In addition, we can use a cyclic cross-correlation to maintain a constant overlap length for any set of time shifts. This is achieved by creating a periodic summation for each $I_{i}^{j}$ defined as
\begin{equation}
\tilde{I}_i^{j}(t) = \sum_{k = -\infty}^{\infty}{I_i^{j}(t - kT_i^{j})} ,
\label{eq:period}
\end{equation}
where $T_i^{j}$ is the length of time $I_{i}^{j}$ elapses. We then cross-correlate the $N$ time series for each $j$th nucleotide together using
\begin{equation}
\begin{split}
g_{N}^{j}(\tau_1, \ldots, \tau_{s-1}, \tau_{s+1}, &\ldots, \tau_N) = \\
&\frac{1}{T_{s}^{j}} \int_{-\infty}^{\infty}dt\,I_s^{j}(t) \prod_{i \neq s}^{N}{\tilde{I}_i^{j}(t+\tau_i)},
\end{split}
\label{eq:cross}
\end{equation}
to obtain a single function.

The function $g_{N}^{j}$ is the $N$-point cross-correlation, while $\tau_i$ is the time shift of the $i$th electrode pair. $I_s^{j}$ is the time series for the $j$th nucleotide with the smallest length of time, $T_{s}^{j}$. We choose all but $I_s^{j}$ to be periodically extended so that no overlapping current values are included more than once within any $\tilde{I}_i^{j}$. By dividing by $T_{s}^{j}$ the cross-correlation values are normalized to be independent of the time overlap.

Due to the nature of the probability distributions for the tunneling currents (see Fig. \ref{fig:one_pair}), $g_{N}^{j}$ covers several orders of magnitude and thus is best portrayed as ${\rm log}(g_{N}^{j})$, where the log is taken as base 10. We then bin cross-correlation values over the set of $\{\tau_i\}$ such that each distinct point ${\rm log}(g_{N}^{j}(\tau_1, \ldots, \tau_{s-1}, \tau_{s+1}, \ldots, \tau_N))$ with $\tau_i \in (-T_i^{j} / 2 ,\, T_i^{j} / 2]$ is a dimension of the histogram (i.e., including only one period for each $\tau_i$). On the basis of how we have constructed the set $\{I_{i}^{j}\}$ using the properties of statistical independence, we can treat each point in $g_{N}^{j}$, and consequently ${\rm log}(g_{N}^{j})$, as following the same probability distribution. As a result, the joint probability distribution is symmetric over the exchange of any two dimensions. However, because of the built-in correlation between each point of the cross-correlation $g_{N}^{j}$, the joint probability distribution is not purely isotropic and none of the dimensions may be traced out.
%Declaring $l_i^{j}$ as the length of time used by $I_{i}^{j}$, $L^{j} = \min_i{[l_i^{j} / 2 + \tau_i]} - \max_i{[-l_i^{j} / 2 + \tau_i]}$ is the length of the intersection of all of the time series.
%
\begin{figure}
\includegraphics[width=\columnwidth]{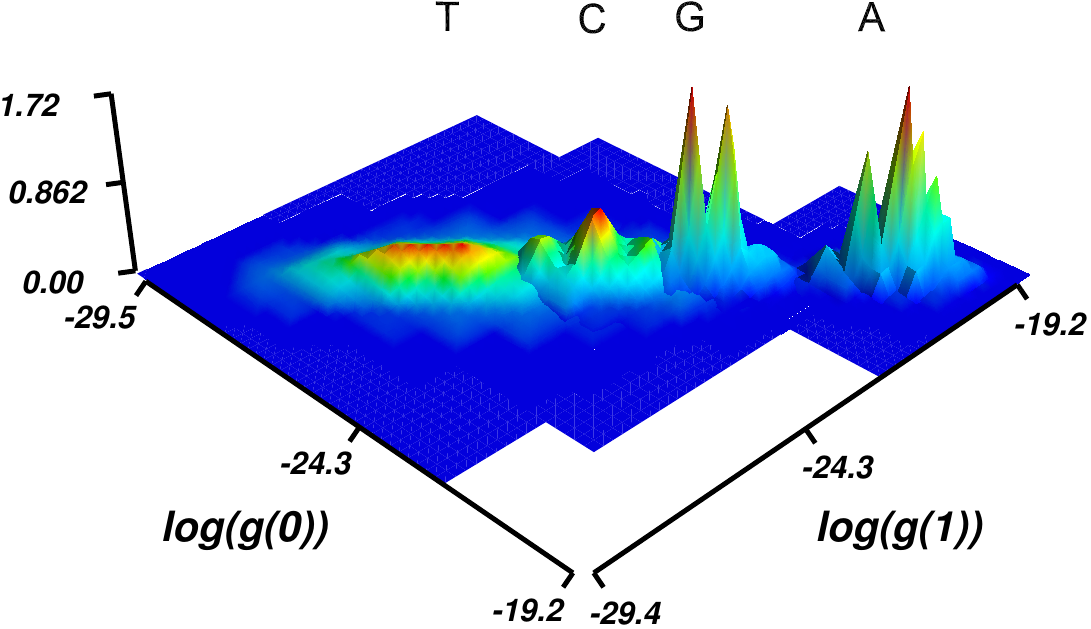}
\caption{\label{fig:joint}(Color online) Normalized joint distributions, $z = P_{2,2}^{j}({\rm log}(g_{2,2}^{j}))$, for $j = A, C, G, T$. Since $g_{2,2}^{j}$ only has $d = 2$ distinct points there are only 2 independent dimensions in the joint distributions. These joint distributions are linear interpolations of the original histograms. The color is only used to further illustrate changes in the $z$-axis and does not represent the same $z$ values across different distributions.}
\end{figure}

For ease of computation we build each $I_{i}^{j}$ to have equal length ($T_i^{j} = T$) and uniform spacing ($\Delta t$) implying that the number of measurements taken at each electrode pair, $m = T / \Delta t$, is the same for each nucleotide. Since the order of the nucleotides does not affect the outcome we just need to compute $g_{N}^{j}$ for $j = A, C, G, T$ to understand how cross-correlating the time series from all electrode pairs together affects the distinguishability of the four DNA bases. However, to gain this understanding we must construct the set $\{I_{i}^{j}\}$ for $N$ electrode pairs with a certain $m$ value and compute $g_{N}^{A, C, G, T}$ many times so that we have a large pool of cross-correlations to interpret and histogram. In this case $g_{N}^{j}$ would have $d = m^{N-1}$ distinct points, meaning that the joint probability distribution for ${\rm log}(g_{N}^{j})$ would be $d$-dimensional. For reference purposes we add the number of measurements per electrode pair, $m$, as an index to the cross-correlation function, now $g_{N,m}^{j}$, and define $g_{N,m}^{j}(k),\, k \in [0,\, d-1]$ as the $k$th point of the cross-correlation function, essentially flattening the set $\{\tau_i\}$ to one index $k$. After creating the histogram for ${\rm log}(g_{N,m}^{j})$ we linearly interpolate it to obtain the continuous joint probability distribution $P_{N,m}^{j}({\rm log}(g_{N,m}^{j}))$, as seen in Fig. \ref{fig:joint}.

With $P_{N,m}^{j}$ for $j = A, C, G, T$ determined with a given number of pairs of electrodes, $N$, and measurements per pair, $m$, we can now compute the distinguishability of the DNA bases. To do this we calculate the average probability of incorrectly determining the identity of a DNA base given a set of tunneling current time series, $\{I_{i}^{j}\}$, from the corresponding nucleotide. This can be expressed by the following equation as
\begin{equation}
\begin{split}
e_{N,m}^{X} &= \left\langle \frac{\sum_{j \neq X}{\tilde{P}_{N,m}^{j}(g_{N,m}^{X})}}{\sum_{j = A,C,G,T}{\tilde{P}_{N,m}^{j}(g_{N,m}^{X})}} \right\rangle _{g_{N,m}^{X}} \\
          &= \left\langle \frac{\sum_{j \neq X}{P_{N,m}^{j}({\rm log}(g_{N,m}^{X}))}}{\sum_{j = A,C,G,T}{P_{N,m}^{j}({\rm log}(g_{N,m}^{X}))}} \right\rangle _{g_{N,m}^{X}} ,
\end{split}
\label{eq:error}
\end{equation}
where $e_{N,m}^{X}$ is the error probability of choosing base $X$ correctly with $N$ pairs of electrodes and $m$ measurements per pair while $\tilde{P}_{N,m}^{j}$ is the probability distribution for $g_{N,m}^{j}$ instead of ${\rm log}(g_{N,m}^{j})$. The average is an ensemble average taken over all possible cross-correlation functions for base $X$. Then we average $e_{N,m}^{X}$ over all of the DNA bases, $X = A, C, G, T$, to obtain the average error probability per base to sequence DNA:
\begin{equation}
E_{N,m} = \frac{1}{4} \sum_{X = A,C,G,T}{e_{N,m}^{X}}.
\label{eq:avg_error}
\end{equation}
\begin{figure}
\includegraphics[width=\columnwidth]{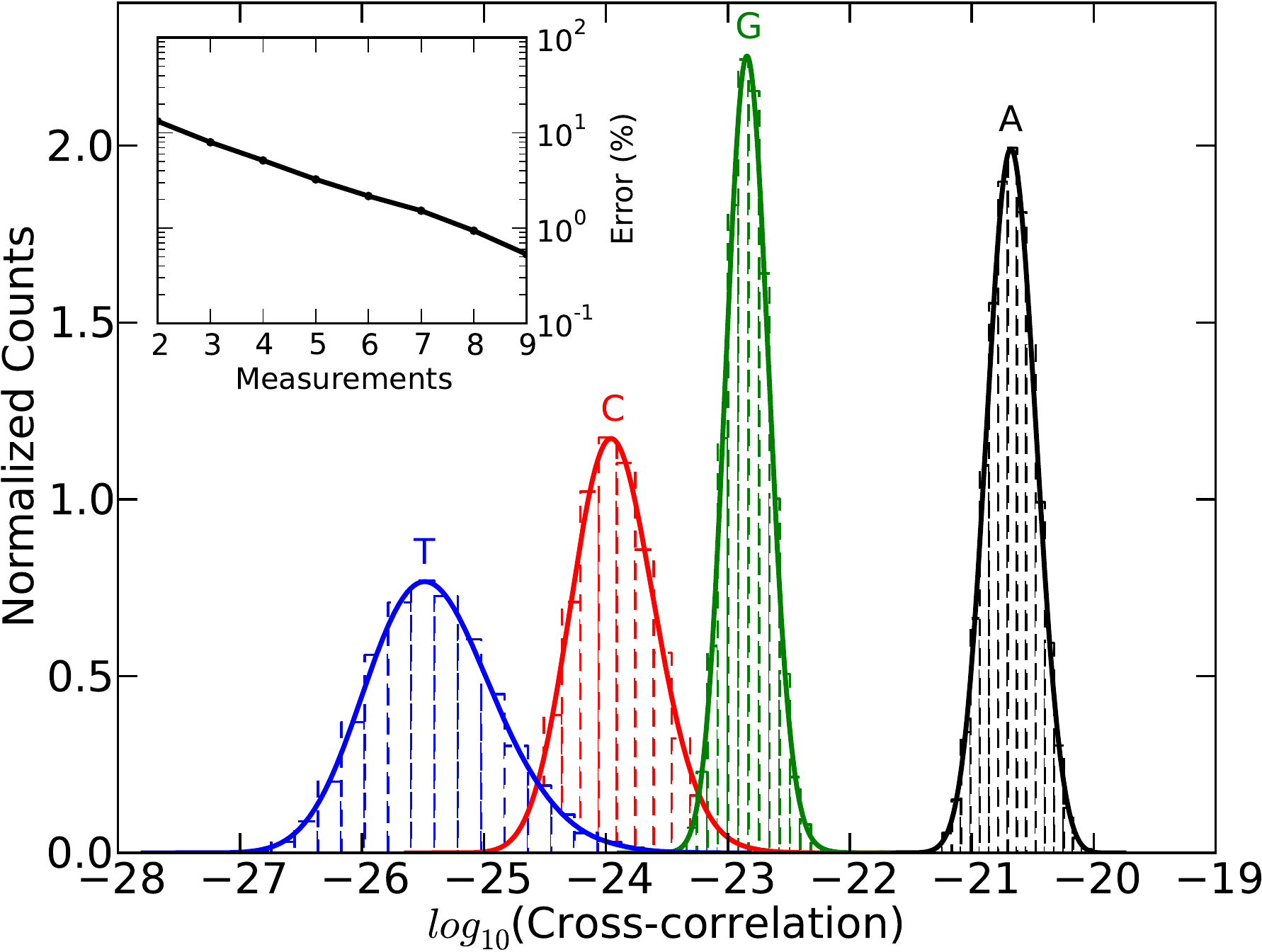}
\includegraphics[width=\columnwidth]{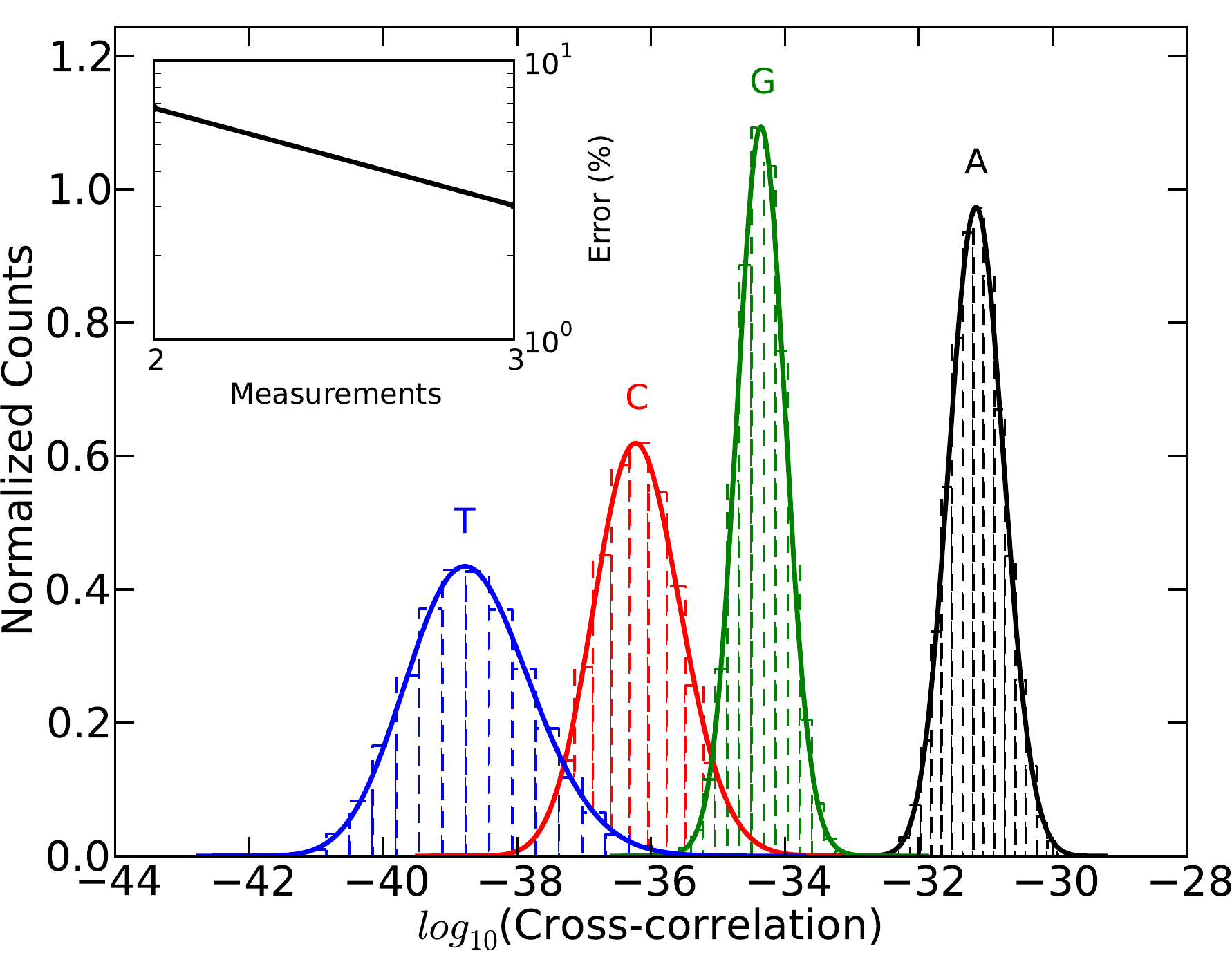}
\caption{\label{fig:multiple_pair}(Color online) Normalized distributions for ${\rm log}(g_{2,9}^{j}(0))$ (top) and ${\rm log}(g_{3,3}^{j}(0))$ (bottom) for $j = A, C, G, T$, where the solid lines are cubic spline mirror-symmetric interpolations of the dashed line histograms. The insets plot the sequencing error percentage per base for $N=2$ (top) and $N=3$ (bottom) on a log scale against the number of measurements per base per electrode pair, $m$.}
\end{figure}
\section{Results and Discussion}
\label{sec:results}

With a collection of error probabilities for different values of $m$ and $N$ we can now evaluate the efficacy of this multiplexing technique. We have calculated $E_{N,m}$ for $N = 2, m = 2 - 9$ and $N = 3, m = 2 - 3$, as illustrated in Fig. \ref{fig:multiple_pair}. For both $N = 2$ and $N = 3$, $E_{N,m}$ decreases linearly with increasing $m$ on a logarithmic scale, meaning $E_{N,m} \sim \beta e^{-a m}$ where $\beta$ and $a$ are positive constants. Due to limited error data for $N = 3$, we compared means and variances to confirm this general trend. Compared to the sequencing error with a single pair of electrodes ($E_{N = 1, m}$), which is also linear with $m$ on a log scale (see the upper inset of Fig. \ref{fig:one_pair}), $E_{N = 2, m}$ and $E_{N = 3, m}$ have nearly double and triple, respectively, the linear rate of decline. Because of the exponential relationship with $m$, we can generously claim
\begin{equation}
E_{N = 2, m} \sim \beta (E_{N = 1, m} / \beta)^2,
\label{eq:error_2}
\end{equation}
and
\begin{equation}
E_{N = 3, m} \sim \beta (E_{N = 1, m} / \beta)^3.
\label{eq:error_3}
\end{equation}
Therefore the improvement in identification errors is significant. In fact, more generally we can assume
\begin{equation}
E_{N, m} \sim \beta (E_{1, m} / \beta)^N.
\label{eq:error_N}
\end{equation}

This result can be easily justified. If the cross-correlation of the $N$ current signals for each base $j$, $\{I_{i}^{j}\}$, from an $N$ electrode pair system did not lose any of the information contained in the original signals, then Eq. \eqref{eq:error_N} would not be generous at all but instead nearly exact. However, a cross-correlation of two different signals certainly results in a loss of information, which manifests itself in the sequencing error by decreasing the exponent $N$ by some factor $\alpha$ representing the fraction of information that was preserved. In other words, the original $\tilde{N}$ signals contain $\tilde{N} \tilde{m}$ points of information, but when cross-correlated what remains is some fraction of that, $\alpha \tilde{N} \tilde{m}$, which results in a more accurate relation between $E_{N,m}$ and $E_{1,m}$,
\begin{equation}
E_{\tilde{N}, \tilde{m}} \sim E_{N = 1, m = \alpha \tilde{N} \tilde{m}} \sim \beta (E_{N = 1, m = \tilde{m}} / \beta)^{\alpha \tilde{N}} .
\label{eq:error_comp}
\end{equation}
By calculating the slope of each line, ${\rm log} \, E_{N,m}$ against $m$ for $N = 2, 3$, with a linear regression we obtain $\alpha = 0.83$ for $N = 2$ and $\alpha = 1.00$ for $N = 3$. This suggests that $\alpha$ saturates to 1 as $N$ increases since with a higher $N$ comes a better chance to reconstruct the original signals from the cross-correlation.

On inspection of Eq. \eqref{eq:error}, one should notice that $e_{N,m}^{X}$ only depends on the probabilities, $P_{N,m}^{j}({\rm log}(g_{N,m}^{X}))$ with $j = A, C, G, T$, and not explicitly on $N$ or $m$. As a result, for the error to decrease as it does for $N = 2, 3$ the joint probability distributions for all 4 bases, $P_{N,m}^{j}$ where $j = A, C, G, T$, must grow farther and farther apart as $N$ or $m$ is increased to reduce their overlap. This is indeed the case and we can study the degree to which the distributions are separated by analyzing the moments of the distributions. Since analyzing the form of the joint distributions, as in Fig. \ref{fig:joint}, becomes too difficult as the number of dimensions, $d = m^{N-1}$, is increased, we settle with analyzing the probability distributions for a single point of the cross-correlation function (e.g., Fig. \ref{fig:multiple_pair}).

Because the distributions in Fig. \ref{fig:multiple_pair} have only one independent variable, ${\rm log}(g_{2,9}^{j}(0))$ for the top and ${\rm log} (g_{3,3}^{j}(0))$ for the bottom, they are fairly smooth due to the integration over all of the other dimensions of the joint distribution. The distributions in Fig. \ref{fig:multiple_pair} are well approximated by normal distributions, which makes the distributions for $g_{2,9}^{j}(0)$, $g_{3,3}^{j}(0)$, and generally any other single point of $g_{N,m}^{j}$ for any $N$ and $m$, approximately log-normal by definition.

A log-normal random variable, $Y$, is best characterized by the mean, $\hat{\mu}$, and variance, $\hat{\sigma}^2$, of ${\rm ln} \, Y$, which follows a normal distribution. ${\rm log} \, Y$ is related to ${\rm ln} \, Y$ with a mean of $\mu = \hat{\mu} / c$ and a variance of $\sigma^2 = \hat{\sigma}^2 / c^2$, where $c = {\rm ln} \, 10$. We can also approximate the original distributions for ${\rm log} \, I_{i}^{j}$ in Fig. \ref{fig:one_pair} as normal, making the distributions for $I_{i}^{j}$ approximately log-normal as well.

If we then examine the discrete form of Eq. \eqref{eq:cross} we find that the unshifted point of the cross-correlation function, labeled $g_{N,m}^{j}(p)$, can be written as
\begin{equation}
g_{N,m}^{j}(p) = \frac{1}{m}\sum_{k = 0}^{m - 1}{\prod_{i = 1}^{N}{\bar{I}_{i}^{j}(k)}} ,
\label{eq:discrete_cross}
\end{equation}
where $\bar{I}_{i}^{j}$ is the discrete form of $I_{i}^{j}$ indexed by measurement number. Any other point in $g_{N,m}^{j}$ is a similar sum of products except that the set of discrete currents has been shifted. The product of any number of log-normal random variables is also log-normal, with its mean and variance parameters defined as the addition of the mean and variance parameters of the random variables that went into the product. Since, for a given base $j$ and any index $k$, every pair of electrodes' current value, $\bar{I}_{i}^{j}(k)$, follows the same probability distribution, the mean and variance parameters for $\prod_{i = 1}^{N}{\bar{I}_{i}^{j}(k)}$ are simply $N\hat{\mu}_1$ and $N\hat{\sigma}_1^2$, respectively. Here, $\hat{\mu}_1$ is the mean of the natural log of the tunneling current with 1 pair of electrodes while $\hat{\sigma}_1^2$ is the variance whereas $\mu_1$ and $\sigma_1^2$ would be the mean and variance of the base 10 log of the tunneling current, as in Fig. \ref{fig:one_pair}. Recalling the properties of independence built-in to the set of $\{I_{i}^{j}\}$, we know that each product in the summation is independent. Therefore we can use the Fenton-Wilkinson approximation, \cite{fenton1960sum}, to obtain the mean and variance of ${\rm log}(g_{N,m}^{j}(0))$ (exactly depicted in Fig. \ref{fig:multiple_pair}) from $\mu_1$ and $\sigma_1^2$,
\begin{equation}
\begin{split}
\sigma_{N,m}^{2} = \frac{\hat{\sigma}_{N,m}^{2}}{c^2} &= \frac{{\rm ln}[1 + (e^{N \hat{\sigma}_1^2} - 1) / m]}{c^2} \\
&= \frac{{\rm ln}[1 + (e^{Nc^2 \sigma_1^2} - 1) / m]}{c^2} ,
\end{split}
\label{eq:fenton_var}
\end{equation}
\begin{equation}
\begin{split}
\mu_{N,m} = \frac{\hat{\mu}_{N,m}}{c} &= \frac{N \hat{\mu}_1 + N \hat{\sigma}_1^2 / 2 - \hat{\sigma}_{N,m}^2 / 2}{c} \\
&= N \mu_1 + cN \sigma_1^2 / 2 - c \sigma_{N,m}^2 / 2 ,
\end{split}
\label{eq:fenton_mean}
\end{equation}
where $\sigma_{N,m}^{2}$ and $\mu_{N,m}$ are the variance and mean of ${\rm log}(g_{N,m}^{j}(0))$ while $\hat{\sigma}_{N,m}^{2}$ and $\hat{\mu}_{N,m}$ are the variance and mean of ${\rm ln}(g_{N,m}^{j}(0))$, for a certain value of $j$. The Fenton-Wilkinson approximation assumes that the sum of log-normal random variables is also log-normal, which is not exact, and then derives the mean and variance parameters by moment matching \cite{fenton1960sum}.

$\mu_{N,m}$ changes dramatically with $N$, but not much with $m$. Therefore as $m$ is increased with a fixed $N$, it is mostly the change in $\sigma_{N,m}^{2}$ that is responsible for the reduced overlap between the cross-correlation distributions and consequently the reduced sequencing error, $E_{N,m}$. While the mean of ${\rm log}(g_{N = 2,m}^{j})$ coincides almost exactly with $\mu_{N = 2,m}$, the variance of ${\rm log}(g_{N = 2,m}^{j})$ can differ from $\sigma_{N = 2,m}^{2}$. In the lower inset of Fig. \ref{fig:one_pair} we plot $\sigma_{N = 2,m}^{2}$ divided by the exact variance of ${\rm log}(g_{N = 2,m}^{j})$ against $m$ for $j = A, C, G, T$ to evaluate the performance of the Fenton-Wilkinson approximation. We can see that all four lines seem to be asymptotically approaching some maximum correction factor. The variance for adenine and guanine is fairly well represented by the approximation, explained by the fact that the log(Current/$A$) distributions for those two bases are closest to resembling normal distributions. Thymine's log(Current/$A$) distribution appears to have a bimodal component, which explains why the Fenton-Wilkinson approximation badly represents the variance of ${\rm log}(g_{N = 2,m}^{T})$. Nevertheless, the approximation can be used as an analytical upper bound on the exact variance of ${\rm log}(g_{N = 2,m}^{j})$ for $j = A, C, G, T$. This variance is an indicator for the sequencing error but it is not sufficient to determine the error alone since the joint distributions are needed.
\section{Conclusions}
\label{sec:conclusions}

An enhancement to the sequencing by tunneling method is proposed, in which $N$ pairs of electrodes are built in series along a synthetic nanochannel. The ssDNA  is forced through the channel using a longitudinal field, as in the original method \cite{zwolak2005electronic,Lagerqvist2006,Krems2009}, and potentially controlled with gate modulation of nanochannel surface charges \cite{he2011controlling}. In this manner the strand of ssDNA passes by each pair of electrodes for long enough to gather a minimum of $m$ tunneling current measurements, where $m$ is determined by the level of sequencing error desired. Each current time series for each base, $I_{i}^{j}$, is then cross-correlated together using a cyclic method to balance the resultant function. With these cross-correlations, one may identify the DNA base by referring to cross-correlation probability distributions that would be obtained from a calibration run.

We have shown that indeed the sequencing error is significantly reduced as the number of pairs of electrodes, $N$, is increased. Compared to the sequencing ability of a single pair of electrodes, cross-correlating $N$ pairs of electrodes is {\it exponentially} better due to the approximately log-normal nature of the original tunneling current probability distributions. We have also used the Fenton-Wilkinson approximation to analytically describe the mean and variance of the cross-correlations that are used to distinguish the DNA bases. When bandwidth is limited, this sequencing method is useful to allow a smaller electrode gap residence time while still promising to consistently identify the DNA bases correctly.
%%%%%%%%%%%%%%%%%%%%%%%%%%%%%%%%%%%%%%%%%%%%%%%%%%%%%%%%%%%%%%%%%%%%%
%% End the main part of the manuscript here.
%\begin{figure*}
%\includegraphics[width=0.75\textwidth]{}
%\caption{}
%\label{fig:}
%\end{figure*}
%
\begin{acknowledgements}
This work was supported in part by the National Institutes of Health, US DOE, and ERC-DM-321031. A. V. Balatsky acknowledges useful conversations with T. Ahmed, J. Haraldsen, T. Kawai, and M. Taniguchi.
\end{acknowledgements}

%\printbibliography % biblatex

\bibliographystyle{ieeetr}
\bibliography{all_bibs}   % name your BibTeX data base

%\begin{thebibliography}{}
% etc
%\end{thebibliography}

\end{document}